\documentclass{article}

\usepackage{arxiv}

\usepackage[utf8]{inputenc} % allow utf-8 input
\usepackage[T1]{fontenc}    % use 8-bit T1 fonts
\usepackage{hyperref}       % hyperlinks
\usepackage{url}            % simple URL typesetting
\usepackage{booktabs}       % professional-quality tables
\usepackage{amsfonts}       % blackboard math symbols
\usepackage{nicefrac}       % compact symbols for 1/2, etc.
\usepackage{microtype}      % microtypography
\usepackage{lipsum}
\usepackage{graphicx}
\graphicspath{ {./images/} }

\title{Repetitive Transcranial magnetic stimulation and epilepsy: A brief essay}

\author{
 Arsalan Heidarpanah \\
  Department of Biomedical Engineering\\
  Islamic Azad University, South Tehran Branch\\
  Tehran, Iran \\
  \texttt{st\_a\_heidarpanah@azad.ac.ir} \\
}

\begin{document}
\maketitle
\begin{abstract}
During the last three decades, many studies have been conducted in the field of treatment with non-invasive methods. In this way, researchers try to use alternative methods including the use of electromagnetic waves in the treatment of diseases. As a result, the therapeutic use of electromagnetic waves in the treatment of neurological diseases has made significant progress. Among the various techniques that have revolutionized the non-invasive treatment of neurological disorders, there is a remarkable technique called Repetitive Transcranial Magnetic Stimulation (rTMS). On the other hand, there is a wide range of neurological conditions (like epilepsy) that are somewhat drug-resistant or can only be controlled with high-risk treatments. In this article, the effect of rTMS on epilepsy is investigated. 
\end{abstract}

% keywords can be removed

\section{Introduction}
Noninvasive brain stimulation is a promising research topic concentrated on clinical neurophysiology, with a wide range of applications from disease diagnosis and therapeautic usage, to pathophysiological study of cortical changes and cortical function mapping.
There are various techniques proposed to stimulate the brain, using electrical and magnetic shocks from the skin through the skull. One of these techniques is the use of a transcranial stimulation device with a repetitive magnetic field used to treat various diseases, including epilepsy. Repetitive transcranial magnetic stimulation (rTMS) technique is a type of TMS that has higher efficiency compared to an ordinary TMS device and has more efficacy in the treatment of various diseases.
These days, research on the control and treatment of epilepsy based on rTMS devices has increased dramatically. In 1831, for the first time, Faraday stated the relation between electric energy and magnetic field. After that, subsequent experiments by scientists showed that using magnetic coils on human head can create a feeling of vertigo and optimism. Also, scientists were trying to induce similar effects through electromagnetic coils placed on the heads of depressed patients. In fact, the ultimate goal was to stimulate brain nervous system [1]. Ultimately, magnetic coils were used to treat depressive mental disorders through the electrical effects of the magnetic field created on the scalp. In 1934, scientists used the Electro Convulsive Therapy (ECT) devices made for clinical analysis of various mood disorders, for the first time. In this way, they tried to identify the brain mechanisms and treat related diseases [5].
The current use of electromagnetic induction for transcranial stimulation dates back to 1985. At that time, Barker et al. invented the first generation of TMS devices in United Kingdom [1]. Barker scientifically proved the effect of magnetic stimulation on the motor cortex of the human brain. rTMS was initially limited to diagnosing motor neuron disorders. Several years later, scientists suggested that applying rTMS to the cortical regions of the brain would produce antidepressant effects. However, the effects of TMS on the frontal area were not considered in the early stages of TMS development. In 2002, the Canadian Public Health Association, as an official organization, recognized the medical outcomes and benefits of rTMS. In addition, other countries (e.g., United States, United Kingdom, Germany, Japan, etc.) have been working on this method for more than two decades. Currently, repetitive transcranial magnetic stimulation devices have been developed in which various parameters (e.g., the ability to increase or decrease magnetic energy in cortical areas) can be modified precisely to produce the optimum therapeutic effects. TMS was developed in 2006 and used to treat depression. Besides TMS, other methods, e.g., TDCS, TES, and ECS were used, each of which had a series of considerable weaknesses. The rTMS device has had a more favorable performance than other methods. It has also proved that rTMS has a better efficacy than TMS in the control and treatment of epilepsy as well as it is more stabe on the human brain. It is worth mentioning that the US Food and Drug Administration (FDA) has also approved this method on October 8, 2008 [6]. 

\section{Epilepsy}
\label{sec:headings}
Epilepsy is a central nervous system disorder in which the activity of nerve cells in the brain is disrupted and as a result it leads to frequent seizures. The cause of epilepsy is low activity of inhibitory neurons and high activity of excitatory neurons. In some cases, epilepsy is caused due to brain damage, brain cancer, drug/alcohol abuse, and other similar reasons. It has estimated that about 1 percent of the world population as well as about 1.5 percent of Iranian population suffer from epilepsy symptoms [2]. 
Epilepsy is divided into two general types:
\textbf{1. Focal seizure:}	If the cause of the seizure is the abnormal activity of an area of the brain, it is called a focal seizure. This type of seizure, itself is divided into two subgroups: simple focal seizure and cognitive focal seizure.
\textbf{2. General seizure:}	Seizures that involve all areas of the brain are called generalized seizures. There are 6 types of generalized seizures: absence seizures, tonic seizures, clonic seizures, myoclonic seizures, atonic seizures, and tonic-clonic seizure.

\section{Epilepsy: Management and treatment}
\label{sec:headings}

\textbf{Drug therapy}

In most patients, seizures can be controlled by taking anticonvulsant drugs (that are called anti-epileptic drugs as well). Taking medicine in some patients reduces the frequency and severity of seizures. Doctors try to return the activity of neurons to their normal state by prescribing these drugs. The common antiepileptic drugs include Clobazam, Depakine, Carbamazepine, Lamotrigine, Phenytoin, etc. Among the side effects of these drugs, it can be mentioned fatigue, dizziness, weight gain, decrease in bone mass density, skin rashes, lack of coordination in movements, difficulty in speaking, and memory impairment.
Since the antiepileptic drugs are usually highly expensive and there are a wide range of well-known pharmaceutical side effects on the vital systems (e.g., liver, kidney, etc.), researchers try to find alternative non-invasive methods.

\textbf{Surgery}

Surgery is often used when the seizure originates in a small and specific area of the brain and does not interfere with vital functions such as speech, type of spoken language, motor functions, vision, and hearing.
Since surgery is a dangerous and highly invasive procedure, there are only a limited number of physicians advised surgery as a suitable therapeutic method. Among its risks, it can be mentioned the high risk of mortality, and impairment in speech ability, hearing, learning or other senses. Due to the sky-high cost of surgery and the rational desire to take less risk, physicians are generally trying to use non-invasive methods to treat epilepsy patients; That is why medical community have turned to using devices with the least negative effects including devices based on the therapeutic effects of electromagnetic waves.

\textbf{Use of electromagnetic waves}

Recently, use of non-invasive techniques including TMS and repetitive TMS has become highly regarded. Because these devices have fairly low risk compared to invasive methods, and their adverse effects are almost negligible according to the studies conducted. Furthermore, their performance is significantly better and they are more effective, compared to other methods [1]. 
TMS may be divided into three main classes: single pulse (spTMS), paired pulse (ppTMS) and repetitive TMS (rTMS). TMS was initially used as a diagnostic tool, but it has gradually become a therapeutic method trusted for using in several psychiatric and neurological disorders. rTMS is a method based on Faraday's law of electromagnetic induction so that the magnetic field outside the brain creates an electric current inside the brain. Therefore, it causes electrical stimulation of the cortex [5]. rTMS is a non-invasive and painless method that is considered highly safe. In addition, low cost and negligible adverse effects are among the other advantages of this method [1].
It is estimated that the total cost of hospitalization and medication for a patient with refractory focal epilepsy in the United States is about 938,800 USD, while the patients are often left untreated despite these huge costs. Meanwhile, after 11 sessions of rTMS treatment with a total cost of 4400 USD (equivalent to 0.5 percent of the hospital costs), the patients remained seizure-free for about 9 months in some cases. In Iran, the cost of treating an epileptic patient (without an insurance coverage) is close to 450.000.000 Rials per year.
Current rTMS equipment is capable of magnetic induction up to two centimeters in the brain. This technology can easily stimulate the areas between the white matter and the gray matter.[1] From this account, the nerve axons transmit the generated currents nearly two centimeters below the coil. In addition, the electrical current that causes changes in nerve activity is in the vicinity of 70 millivolts [2]. rTMS can produce a long-lasting change in neural activity. This sustained effect causes a change in cortical excitability that decreases at low frequency (<1 Hz) and increases at high frequency rTMS (>10 Hz) [3]. In addition, the effects of rTMS can be observed during a treatment using an electroencephalogram device, a non-invasive, easy, and affordable method of recording provides the ability to re-acquire the signal in order to check brain signal changes [4].

\section{rTMS: The mechanism of action}
\label{sec:headings}
The various portions of the TMS mechanism, from the pulse generated by the coil to the intensity of nerve cell stimulation, generally follow laws of Electromanegtics [5]. In general, TMS equipment consists of a transformer to charge a large capacitor, instantaneously discharged to generate a magnetic field pulse in the coil for excitation. In fact, there is a sub-circuit to control temperature, intensity and pulse repetition. The maximum voltage and the generated currents are approximately 2000 volts and 10000 amperes, respectively. To generate a short pulse (about 250 microseconds or 1.4000 seconds) it may be necessary a high voltage electrical current. In the TMS mechanism, when the device is charged with an electric current, magnetic fields occur around the coil. According to Faraday's law, if there are two coils next to each other, one will be energized while another will be de-energized, because the current in the first coil stops and a very short pulse is transmitted to the second one.

\section{Conclusion}
\label{sec:headings}
Repetitive transcranial magnetic stimulation (rTMS) is a non-invasive procedure for a wide range of neurological conditions, including epilepsy that uses electromagnetic waves based on Faraday’s law. Regard to cost consideration, safety and effectiveness, rTMS technique is considered much better theraupeutic approach compared to various alternative invasive and non-invasive methods.

\section{References}
\label{sec:headings}
1- History, Studies and Specific Uses of Repetitive Transcranial Magnetic Stimulation (RTMS) in Treating Epilepsy. Noohi S, Amirsalari S. Iran J Child Neurol. Winter 2016; 10(1):1-8.

2- Perampanel for tonic-clonic seizures in idiopathic generalized epilepsy A randomized trial. Jacqueline A. French, MD,  Gregory L. Krauss, MD, Robert T. Wechsler, MD, PhD, Xue-Feng Wang, MD, Bree DiVentura, MBA, Christian Brandt, MD, Eugen Trinka, MD, MSc, Terence J. O'Brien, MD, BS, Antonio Laurenza, MD, Anna Patten, PhD, and Francesco Bibbiani, MD. Neurology. 2015 Sep 15; 85(11): 950–957.

3- Safety and tolerability of repetitive transcranial magnetic stimulation in patients with epilepsy: a review of the literature. Erica Hyunji Bae a, Lara M. Schrader b, Katsuyuki Machii c, Miguel Alonso-Alonso c, James J. Riviello Jr. a, Alvaro Pascual-Leone c, Alexander Rotenberg a,c,*. Epilepsy \& Behavior 10 (2007) 521–528.

4- Detection of early stage Alzheimer’s disease using EEG relative power with deep neural network. D Kim, K Kim - 2018 40th Annual International Conference of..,2018 - ieeexplore.ieee.org.

5- Rudorfer, M. V., Henry, M. E., \& Sackheim, H. A. (1997). Electroconvulsive therapy. In A. Tasman, J. \& J. A Lieberman (Eds.), Psychiatry (1535-1556)

6- Connolly KR, Helmer A, Cristancho MA, Cristancho P, John PO. Effectiveness of transcranial magnetic stimulation in clinical practice post-FDA approval in the United States: results observed with the first 100 consecutive cases of depression at an academic medical center. The Journal of clinical psychiatry. 2012 Apr 15;73(4):5611.

\end{document}